# The time-evolving impact of tree size on nighttime street canyon microclimate: Wind tunnel modeling of aerodynamic effects and heat removal

Yongling Zhao [a, *], Haiwei Li [a, b], Ronita Bardhan [b], Aytac Kubilay [a], Qi Li [c], Jan Carmeliet [a]

[a] Department of Mechanical and Process Engineering, ETH Zürich, Zürich, Switzerland
[b] The Sustainable Design Group, Department of Architecture, University of Cambridge, Cambridge, UK
[c] School of Civil and Environmental Engineering, Cornell University, Ithaca, USA

*Email: yozhao@ethz.ch

**Abstract**

Urban trees play a crucial role in urban climate in many aspects. However, existing research has not adequately explored the impact from a time-evolving perspective, that is, tree growth over time. To bridge this research gap, this study investigates in a wind tunnel the effects of tree-to-canyon foliage cover and relative height (0.32~1.1 times canyon height), mimicking growth of trees, on conditions in street canyons during moderate and extreme heat. The results reveal that trees may affect canyon-wide ventilation and heat removal in two different scenarios. First, when canyons are in isothermal conditions, medium and large trees, that fill half the canyon height or reach slightly above the canyon, decelerate the shear layer and weaken the vortical flow, as a result reducing the canyon-wide ventilation. Second, in extreme heat conditions, medium and large trees trap heat at the pedestrian level due to the blockage of air entrainment and the suppression of upward buoyancy-driven flow from the ground surface. An air temperature rise that corresponds to 1.5°C in a full-scale urban setting is observed in measurements. These observations suggest that urban trees' foliage cover must be managed for a canyon's optimal ventilation and heat removal during nighttime.

# 1. Introduction

Urban dwellers are suffering from global climate change and various urban environmental issues, including urban heat island (UHI) (Fan, Hunt, et al., 2021; Grimm et al., 2008; Kong et al., 2021; Moonen et al., 2012; Peng et al., 2022; Zhao et al., 2014), air pollution (Michetti et al., 2022) and heat burden (Aghamolaei et al., 2021). Urban green infrastructure (UGI) is usually promoted as a versatile measure in urban planning, improving urban sustainability and livability (Bowler et al., 2010; Pan et al., 2022; Pataki et al., 2011; Ricci et al., 2022; Susca, 2019). The use of street trees is one of the most common measures in cities, which may play positive roles in mitigating urban heat (Kubilay et al., 2020), reducing building cooling demand (Ng et al., 2012; Wang et al., 2016), improving air quality (Huang et al., 2019; Jim et al., 2008; Willis et al., 2017), and also adding cultural aesthetic values to cities (Locke et al., 2015).

Trees in street canyon are seen to modify airflow around their crowns, which involves complex airflow-leaves interaction (Fan, Ge, et al., 2021). Effects of trees have been investigated by wind tunnel measurements (Fellini et al., 2022; Gromke, 2011; Manickathan et al., 2018a; Zhao et al., 2023), field measurements (Armson et al., 2013; Gillner et al., 2015; Shashua-Bar et al., 2009) and Computational Fluid Dynamics (CFD) simulations (Manickathan et al., 2018b). Gromke (2011) applied a novel concept in wind tunnel modeling of street trees using fiber-like porous material to mimic artifical leaves. The porous material is characterised by two main properties, pore volume fraction and pressure loss coefficient. Manickathan et al. (2018a) recently compared the drag coefficient and other aerodynamic characteristics between real and artificial trees in the wind tunnel. The results showed that the artificial and real leaves present similar aerodynamic characteristics. A most recent work of Fellini et al. (2022) examined the impact of tree density on street canyon pollutant concentration and ventilation. The wind tunnel measurements reveal that the average level of pollution and overall ventilation efficiency are not significantly affected by the presence of the trees that are lower than the height of surrounding buildings.

The cooling effects of urban street trees on microclimate in seasonal and diurnal cycles in different climates have been studied extensively (Jiao et al., 2017; Jonsson, 2004; Meili, Acero, et al., 2021; Peng et al., 2022; Rahman et al., 2020; Wang et al., 2019; Ziter et al., 2019). The effectiveness of cooling achieved by urban trees varies seasonally as multiple factors may apply, including air humidity, tree-to-canyon foliage cover ratio, species, moisture, and albedo (Jonsson, 2004; Wang et al., 2021). In a study by Meili, Acero, et al. (2021), simulation results from an ecohydrological model shows that trees provide a significant cooling effect during the warm season. However, a negligible effect or even slight warming are seen in cool seasons with pronounced seasonal temperature variations, for instance in Phoenix, Melbourne, and Zurich. Some studies have shown an air temperature increase up to 0.6 °C at night (Konarska et al., 2016; Rahman et al., 2017; Shashua-Bar et al., 2009).

Gillner et al. (2015) studied the diurnal cooling potential of six tree species on three hot days in Dresden, Germany. During the daytime, the local air temperature is reduced by 0.8°C to 2.2°C. The cooling effects vary, depending on different background climates and soil conditions, dominant plant species, plant morphologies, and plant densities. Another study in Manchester, United Kingdom, reported that the surface temperature is reduced by an average of 12°C, depending on the tree species and leaf area index (LAI) (Armson et al., 2013).

The cooling mechanisms, aerodynamic effects, longwave and shortwave radiation blockage, and evapotranspiration of leaves jointly determine air and surface temperatures, and wind flow conditions (Mughal et al., 2021). In terms of modelling the aerodynamic influence of street trees in CFD simulations, the foliage of trees can be treated as a porous medium, which maintains a high surface-to-volume ratio and skin friction, leading to large drag force or aerodynamic resistance around leaves (Frank et al., 2008; Gross, 1987; Wang et al., 2018; Zölch et al., 2019). The local wind speed is reduced, especially around leave foliage, resulting in a reduction in air ventilation and heat removal in street canyons (Li et al., 2022; Li & Wang, 2018).

Another important aspect is the counteracting radiation blockage effects of trees during a diurnal cycle. Trees provide shading during the daytime, reducing the shortwave solar irradiation at the ground and building surfaces, being one of the main contributors to cooling effects, which has been studied by many researchers (e.g., Alavipanah et al., 2015; Morakinyo et al., 2016; Shashua-Bar et al., 2002; Yu et al., 2020). However, trees also reduce the sky view factor in street canyons (Li & Ratti, 2018), leading to an entrapment of longwave radiation from the ground and building surfaces during night, which may negatively affect the cooling effects at night (Mehrotra et al., 2019; Meili, Manoli, et al., 2021; Song et al., 2015).

Tree transpiration and transpirative cooling vary throughout the day, reaching a high level in the daytime due to the full stomatal opening. However, in darkness or extremely dry weather, trees are found to maintain a minimized transpiration due to the stomatal closure (Oren et al., 1999). Soil water availability for stomatal conductance has also significant impacts on transpiration (Pace et al., 2021). These mechanisms collectively interact and usually lead to daytime cooling and nighttime neutral-to-warming effects. The morphology of the trees is one of the most important factors that heavily influence the interacting mechanisms in many ways. First, the implementation of trees influences the development of urban canopy layer (UCL) by altering the aerodynamic characteristics, e.g. the zero-plane displacement ($z_d$) and aerodynamic roughness length ($z_0$) (Finnigan, 2000; Kent et al., 2017). A study showed that the wind speed could be lowered by up to a factor of three with the dense vegetation reaching higher than the urban canopy (Kent et al., 2017). Second, the relative size of the tree crowns to the canyon opening determines the sky-view factor (SVF), significantly influencing solar irradiation and long-wave radiation exchange, and thus heat absorption and heat release. A field study in a tropical urban complex in Singapore reported that mature street trees might lead to warmer nighttime air temperatures due to trapping of long-wave heat release from urban surfaces (Hien et al., 2010).

The time-evolving impact of trees, due to the growth, on street canyon microclimate could be complex. As time passes, the height and foliage cover of street trees may vary dramatically, which will in turn affect the surrounding airflow dynamics, ventilation and heat removal. However, existing literature considers mainly trees of a given height and foliage cover. Understanding of the time-evolving impact of street trees on urban microclimate is strongly needed. In this study, Planar Particle Image Velocimetry (PIV), thermal infrared imaging, and in-house designed thermocouples are used to measure the air velocity fields, tree surface temperatures, and air temperature fields in the cases of three different tree sizes, respectively. The ground and building surfaces in the domain of interest are heated to a series of temperatures to simulate different thermal conditions, which has been applied in a previous wind tunnel study to simulate buoyant flows in canyons (Zhao et al., 2021). The artificial filament/fiber-like synthetic wadding materials are used to mimick leaves of tree models. The same material has been applied in previous similar wind tunnel studies for modeling tree leaves and vegetation on building facades (Gromke, 2011; Li et al., 2022). Street trees of three different sizes are placed in two rows in consecutive flat and steep canyons. The aerodynamic influences and the heat trapping at nighttime are the primary mechanisms of interest in this study, while biological features, such as low level of transpiration at nighttime, are not considered in the measurements.

The experimental setup of wind tunnel, including PIV, thermocouples and infrared cameras, is described in section 2. The results and analysis regarding the airflow structures, roof-level ventilation, momentum fluxes, thermal conditions at pedestrian-level and leaf heat storage are elaborated in section 3. A summary of the flow structures and the limitation of this work are presented in section 4. Lastly, the concluding remarks are given in section 5.

## 2. Experimental setup
### 2.1 Street canyons and trees

The neighborhood model used in the ETHZ/Empa Atmospheric Boundary Layer wind tunnel is shown in Fig.1. The model is based on a reduced-scale ($M = 1:160$) representation of Glattpark in Zurich, Switzerland (Tsalicoglou et al., 2020). The measurements are carried out on two consecutive street canyons, a first one with buildings of similar height on leeward and windward (referred to as *flat canyon*, FC), and a second one with a higher building windward (referred to as *steep canyon*, SC). Previous measurements from Li et al. (2022) and Zhao et al. (2021) have used the same neighborhood morphology to understand street canyon flows in non-isothermal conditions.

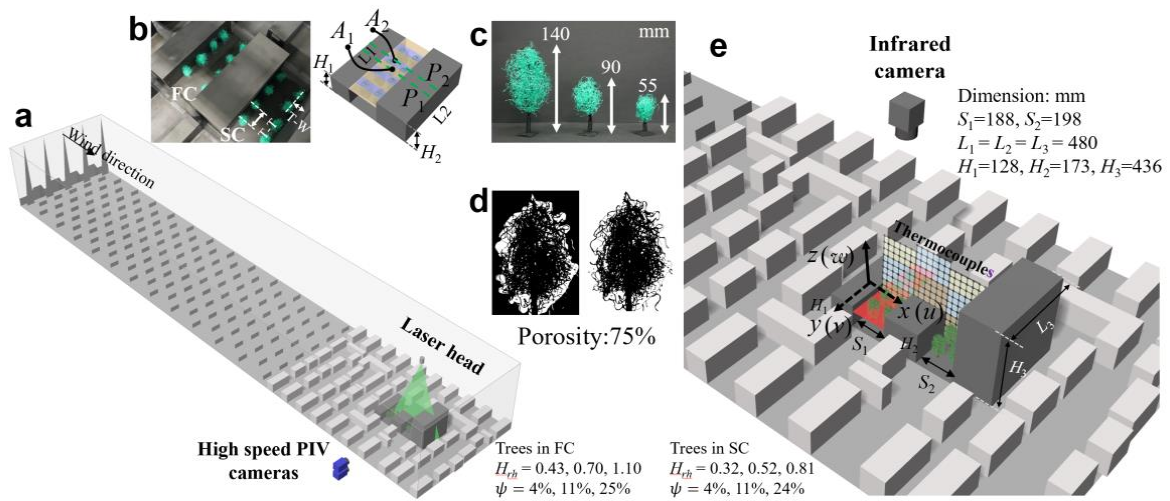

**Fig.1** Experimental setup showing: (a) upstream roughness elements and two street canyons with PIV system, (b) placement of trees and indication of two measurement planes, (c) tree models, (d) determination of coverage of leaves, and (e) infrared imagery and thermocouple arrangement. The setup of the neighbourhood models follows the previous wind tunnel studies of Li et al. (2022) and Zhao et al. (2021).

The heights of trees are chosen to be 8.8 m, 14.4 m, and 22.4 m in full scale. To characterize the height of tree relative to the canyon height, a tree-to-canyon relative height ($H_{rh}$) is proposed and calculated using Eq.(1):

$$H_{rh} = H_{tr}/H_{ca} \qquad (1)$$

where $H_{tr}$ denotes the height of the trees and $H_{ca}$ denotes the height of the canyon.

The calculation above uses the height of the upstream building of the canyon as the canyon height ($H_{ca}$), that is, $H_1$ for FC and $H_2$ for SC. The relative height of trees ($H_{rh}$) at reduced scale is determined to be 0.43, 0.70, and 1.10 for the flat canyon and 0.32, 0.52, and 0.81 for the steep canyon. The selected trees span a wide range of relative heights in measurments.

A tree-to-canyon foliage cover ($\psi$) is also proposed to characterize the percentage of the bulk area where the airflow and heat removal are mostly affected by the presence of trees. The foliage cover ($\psi$) for a canyon is calculated using Eq.(2):

$$\psi = A_{tr}/A_{ca} \qquad (2)$$

where $A_{tr}$ denotes the total ground surface covered by the trees in one canyon and $A_{ca}$ denotes the total area of the ground surface of the corresponding canyon.

Two rows of four trees are equally spaced in each street canyon. The tree-to-canyon foliage cover $\psi$ is approximately determined to be 4%, 11%, and 25% for both flat and steep canyons. Artificial filament/fiber-like synthetic wadding material is used to model the leaves of trees (Gromke, 2011). The coverage ratio of leaves and pore volume fraction are two important factors to indicate the porosity and aerodynamic effect of the leaves (Li et al., 2022), which is determined by calculating the ratio of the tree structure area to the total area of trees in a binary image of the vegetation sample, as shown in Fig.1d. This technique of calculating the coverage ratio of vegetation follows the procedure in Chu et al. (2017). The coverage ratio of the trees in the study is around 0.72-0.78. The crown size, porosity of the tree models, their spacing, and foliage cover are documented in **Appendix A**. The choice of the length scale of the flat and steep canyons is discussed in Zhao et al. (2021) and Li et al. (2022). Shading and transpirative cooling by trees are not modeled and the experimental results should be interpreted as the primary effects of trees during nighttime.

**2.2 Wind and buoyancy conditions**

The profiles of the approaching urban boundary layer used in this series of measurements are determined based on earlier wind tunnel studies (Li et al., 2022), which are also presented in Fig.2. The flat and steep street canyons (denoted as FC and SC in Fig. 1b) are located in the urban boundary layer. In the measurements the approaching free-stream wind speed is at 0.6 m/s at the height of two times of the tall building of the steep canyon, which corresponds to a full-scale gentle breeze at a speed of 4.2 m/s. The estimation of the Reynolds number (*Re*) for the isothermal cases and the Richardson number (*Ri*) for the non-isothermal cases are shown in Table 1.

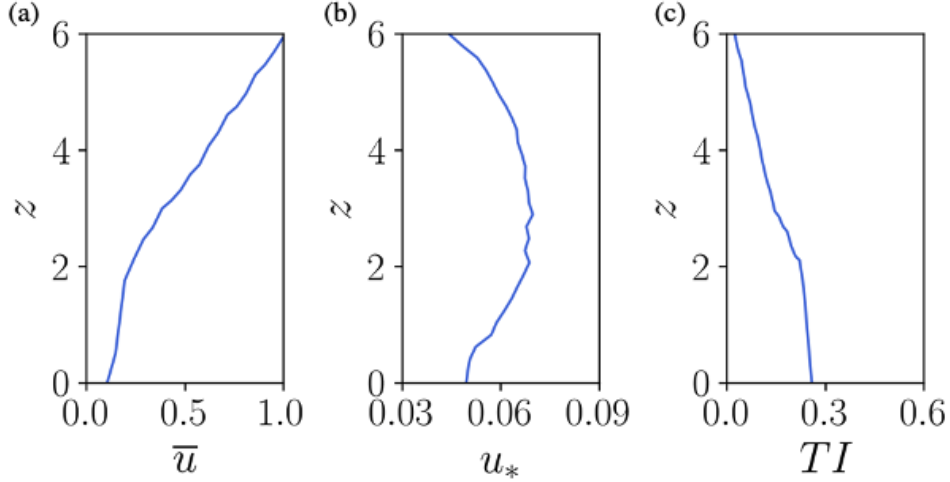

**Fig.2** Time-averaged profiles of (a) streamwise velocity component, (b) friction velocity and (c) turbulent intensity measured at the location of the steep canyon without the presence of building models. The velocity $\bar{u}$ is normalized by the wind tunnel reference velocity $U_{ref\_wt}$ and $z$ is normalized by the $H_1$.

For the intermediate and extreme heat scenarios, the building surface temperature is set at 42°C and 82°C, and the air temperature of the approaching boundary layer flow in measurements is at 20°C. The temperature difference between canyon surfaces (i.e., building and ground surfaces) and the approaching boundary layer applied in measurements corresponds to the temperature difference of 7°C and 19°C in full-scale settings. This level of temperature difference is representative and observed in field measurements (Santamouris et al., 1999). The isothermal or non-isothermal airflow around the flat and steep canyons is characterized by the Reynolds number of local Richardson number, as shown in Eqs. (3-4):

$$Ri_f = \frac{g\beta(T_{ca-wt}-T_{air-wt})H_1}{U_{ref-wt}^2}, \quad Re_f = \frac{U_{ref-wt}}{v}H_1 \quad (3)$$

$$Ri_s = \frac{g\beta(T_{ca-wt}-T_{air-wt})H_2}{U_{ref-wt}^2}, \quad Re_s = \frac{U_{ref-wt}}{v}H_2 \quad (4)$$

Table 1. Model configurations and airflow conditions at the wind tunnel scale and corresponding full scale. $H_{tree\_fs}$ denotes the height of full-scale trees, $T_{ca\_fs}$ the surface temperature of the canyon surfaces (three buildings and the ground in between) and $T_{air\_fs}$ the air temperature of the ambient airflow at full scale, $U_{ref\_fs}$ the reference wind speed at full scale. The subscript '*fs*' denotes full scale quantities and '*wt*' denotes wind tunnel scale quantities.

| Run | Full scale | | | Wind tunnel scale | | | Characteristic numbers | | |
|---|---|---|---|---|---|---|---|---|---|
| | $H_{tree\_fs}$ (m) | $T_{ca\_fs}$ - $T_{air\_fs}$ (°C) | $U_{ref\_fs}$ (m/s) | $H_{tree\_wt}$ (mm) | $T_{ca\_wt}$ - $T_{air-wt}$ (°C) | $U_{ref-wt}$ (m/s) | $Ri_f$ | $Ri_s$ | $Re_f$, $Re_s$ |
| 1 | 22.4 | 0 | 4.2 | 140 | 0 | 0.6 | / | / | 5050, 6830 |
| 2 | 22.4 | 7 | 4.2 | 140 | 22 | 0.6 | $Ri_f$=0.28 | $Ri_s$=0.38 | Relax. |
| 3 | 22.4 | 19 | 4.2 | 140 | 62 | 0.6 | $Ri_f$=0.80 | $Ri_s$=1.08 | Relax. |
| 4 | 14.4 | 0 | 4.2 | 90 | 0 | 0.6 | / | / | 5050, 6830 |
| 5 | 14.4 | 7 | 4.2 | 90 | 22 | 0.6 | $Ri_f$=0.28 | $Ri_s$=0.38 | Relax. |
| 6 | 14.4 | 19 | 4.2 | 90 | 62 | 0.6 | $Ri_f$=0.80 | $Ri_s$=1.08 | Relax. |
| 7 | 8.8 | 0 | 4.2 | 55 | 0 | 0.6 | / | / | 5050, 6830 |
| 8 | 8.8 | 7 | 4.2 | 55 | 22 | 0.6 | $Ri_f$=0.28 | $Ri_s$=0.38 | Relax. |
| 9 | 8.8 | 19 | 4.2 | 55 | 62 | 0.6 | $Ri_f$=0.80 | $Ri_s$=1.08 | Relax. |
| 10 | / | 0 | 4.2 | / | 0 | 0.6 | / | / | 5050, 6830 |
| 11 | / | 7 | 4.2 | / | 22 | 0.6 | $Ri_f$=0.28 | $Ri_s$=0.38 | Relax. |
| 12 | / | 19 | 4.2 | / | 62 | 0.6 | $Ri_f$=0.80 | $Ri_s$=1.08 | Relax. |

where $Ri_f$ and $Re_f$ are the Richardson number and the Reynolds number for the flat canyon, and $Ri_s$ and $Re_s$ are the Richardson number and the Reynolds number for the steep canyon. The steepness ratio for the flat and steep canyon is determined to be 1.35 and 2.52 (Zhao et al., 2021). Here $U_{ref-wt}$ is the wind tunnel incoming flow velocity; $T_{air-wt}$ denotes the wind tunnel upstream airflow temperature; $T_{ca-wt}$ is the surface temperature of the canyons (buildings and ground surfaces) of interest; $H_1$ and $H_2$ are the heights of the leeward building model in the flat and steep canyon; $g$ is the gravitational acceleration; and $β$ is the thermal expansion coefficient. It is worth noting that the Reynolds number is relaxed in the cases where the flow is largely driven by buoyancy, which is discussed in Zhao et al. (2020).

**2.3 PIV, infrared camera and thermocouple temperature measurements**

The measurements involve flow field, airflow temperature and leaf surface average temperature, which are conducted using PIV, thermocouples and infrared imagery as shown in Fig.1. In the experiments, PIV measurements are conducted first without the presence of the thermocouples and the infrared camera. After that, air temperature and leaf surface temperature measurements are conducted using thermocouples and an infrared camera, respectively.

The setup of the PIV measurements is similar to the study by Zhao et al. (2021), where a 532 nm Nd:YAG laser of 200 mJ/pulse is used and two CMOS 12 bit dual-frame high-speed cameras are arranged in a vertical tower configuration to expand the field-of-view (FOV) (Zhao et al. 2019). To obtain the full flow field around the FC and SC, the measurement is conducted for six spanwise-centered planes, the procedure of which is similar to the study by Li et al. (2022). Additional PIV measurements are also performed at planes intersecting crowns of trees to obtain flow fields that are mostly affected by trees. The particle image pairs recorded 200 seconds at the acquisition frequency of 15 Hz, which are post-processed using a 2-pass cross-correlation with an interrogation window from 64 × 64 pixels to 32 × 32 pixels. The average uncertainty is $10^{-3}$ m/s, which is derived from correlation statistics calculated in the post-processing.

The air temperature measurements are conducted using a setup similar to that reported in Li et al. (2022). Due to the large measurement field, the temperature measurements are conducted in several separate runs, and the full temperature field is obtained by stitching them. For each run of 200 seconds, 35 thermocouples are used, giving an approximate spatial resolution of 33 mm × 40 mm in the lower part of the canyons.

The infrared camera measurement is conducted as the last step to measure leaf surface average temperature. In the measurements, the infrared camera is placed about 50 cm above the trees using a tripod and the recordings are controlled remotely outside the wind tunnel. Surface temperature calibrations are performed for the same focal distance of the infrared camera before the post-processing of the infrared images. The uncertainty of the measurement is determined to be 2% of the temperature range adopted in experiments.

## 3. Results
### 3.1 Impact of tree sizes and urban heat on airflow

Fig.3 and Fig.4 present the time-averaged flow fields of streamwise and vertical velocity from PIV measurements with trees of different foliage cover and relative height. The contours represent the magnitude of velocity component (*u* or *w*) and the arrows represent the velocity vector. The impact of trees on street canyon mean flow in the vertical center plane in isothermal conditions (no heat) is examined first, as shown in Fig.3. It is seen in Fig.3a$_1$-a$_4$ that the trees reduce the strength of the vortical flow in both flat and steep canyons. This effect becomes much more apparent with increasing tree size relative to the canyon's height. For the flat canyon, the canyon vortex does not form when the trees reach slightly above the canyon mid-height (Fig.3a$_3$-a$_4$). The trees in the steep canyon affect the vortical flow less, since the strong downward flush flow facilitates the development of the canyon vortex. As the trees grow approximately to the height of the leeward building of the steep canyon, the vortical flow does not establish anymore due to important aerodynamic resistance of tree leaves on the flow.

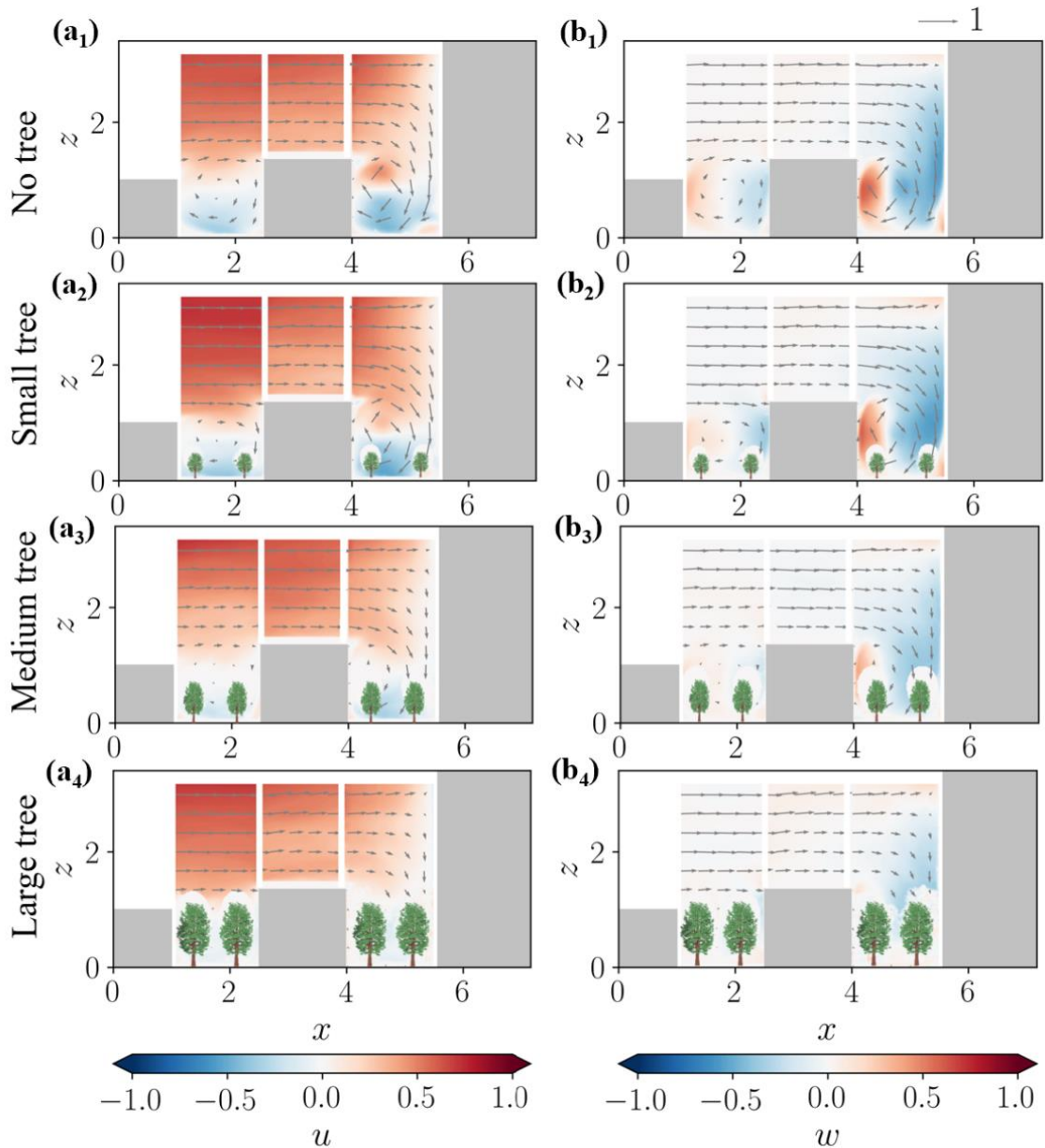

**Fig.3** Contours of (a$_1$-a$_4$) streamwise and (b$_1$-b$_4$) vertical velocity components superimposed with velocity vectors for the flat and steep canyons without and with trees of different foliage cover and relative height in isothermal condition.

In the moderate heat case (Fig4.b$_1$-b$_3$), the airflow in the flat street canyon around the rooftop is primarily buoyancy-driven. Larger trees weaken the upward airflow around the tree's crown, reducing the canyon-wide ventilation and heat removal from the flat canyon.

Moreover, as will be shown below, porous tree crowns may stabilize the airflow due to drag. In the steep canyon for the moderate heat case, a weak vortical flow interacts in the canyon with buoyancy-driven flow, but, a distinct vortical flow can still form due to the strong downward flush flow.

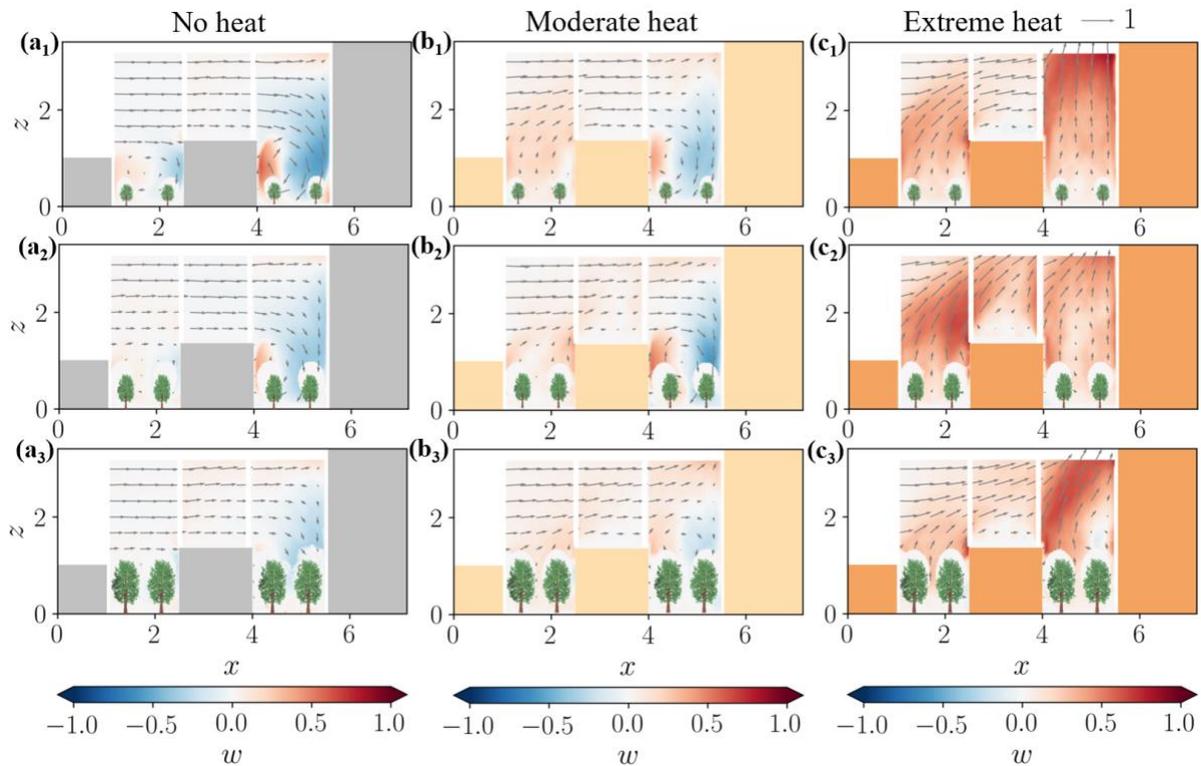

**Fig.4** Contours of vertical velocity components superimposed with velocity vectors for isothermal canyons ($a_1$-$a_4$), canyons subjected to moderate heat ($b_1$-$b_4$), and canyons subjected to extreme heat ($c_1$-$c_4$) in the scenarios with trees of different foliage cover and relative height.

In the case of extreme heat, the airflow in the flat canyon becomes more driven by buoyancy and in the steep canyon is dominated by buoyancy, resulting in a predominantly upwards airflow reaching higher than the street canyon rooftop, as shown in Fig.4$c_1$-$c_3$.

**3.2 Impact on roof-level ventilation**

The impact of trees on the volumetric ventilation rate at roof level is calculated for cases with different tree size subjected to no heat, moderate and extreme heat conditions, as shown in Fig.5. The dimensionless ventilation rate ($Q'$) is comprised of the convective ventilation rate ($Q'_c$) and turbulent ventilation rate ($Q'_t$), as shown in Eqs. (5-7) (Hang et al., 2010; Zhao et al., 2021):

$$Q'_c = \frac{t_r}{V_{c1}} \int_{A_1} \overline{W}_1 \, dA_1 + \frac{t_r}{V_{c2}} \int_{A_2} \overline{W}_2 \, dA_2, \tag{5}$$

$$Q'_t = \pm \frac{t_r}{V_c} \int_A 0.5\sigma \, dA \tag{6}$$

$$Q' = Q'_c + Q'_t. \tag{7}$$

Here, $\overline{W}_1$ and $\overline{W}_2$ denote roof-level time-averaged vertical velocities obtained at measurement planes along a line located at respective canyon spanwise center without trees ($P_1$) and at the intersection ($P_2$) with trees; $\sigma$ the standard deviation of time-varying vertical velocity; $A_1$ and $A_2$ the two areas of the canyon opening at roof level with and without a tree underneath, and $V_{c1}$ and $V_{c2}$ the two corresponding volumes of the canyon. When trees reach above the canyon, the ventilation through tree leaves is not taken into account. A negative $Q'$ suggests the airflow at the canyon opening is dominated by entrainment into the canyon, while a positive $Q'$ implies air removal from the canyon to the roof level.

The time scale $t_r$ that denotes an estimation of the time for the airflow to complete one canyon-wide circulation is calculated using Eq. (8):

$$t_r = 2(H + S)/(U_{ref-wt}\, 2/3) \tag{8}$$

where $H$ denotes the height of the leeward building of the corresponding canyon, $S$ denotes the spacing of the two buildings of the canyon. The reference velocity is chosen to be 2/3 of the freestream approaching flow speed according to the approximation reported in Nakamura et al. (1988). This time scale is used in follow analyses where time series of quantities are involved.

The volumetric ventilation rate approximated by spanwise integration for isothermal case remains at very low magnitude for all cases, as shown in Fig.5a$_1$-a$_4$. The entrainment (dotted lines) and removal (dashed lines) at the roof-level opening approximately balance out each other. A closer examination of the ventilation rate in the cases with medium and large trees reveals that the presence of trees reduces both entrainment and removal, which is due to the weakened vortical flow, as shown in Fig. 3b$_3$-b$_4$. In the case with large trees, there is hardly any canyon-wide vortical flow and thus the volumetric ventilation is dramatically reduced.

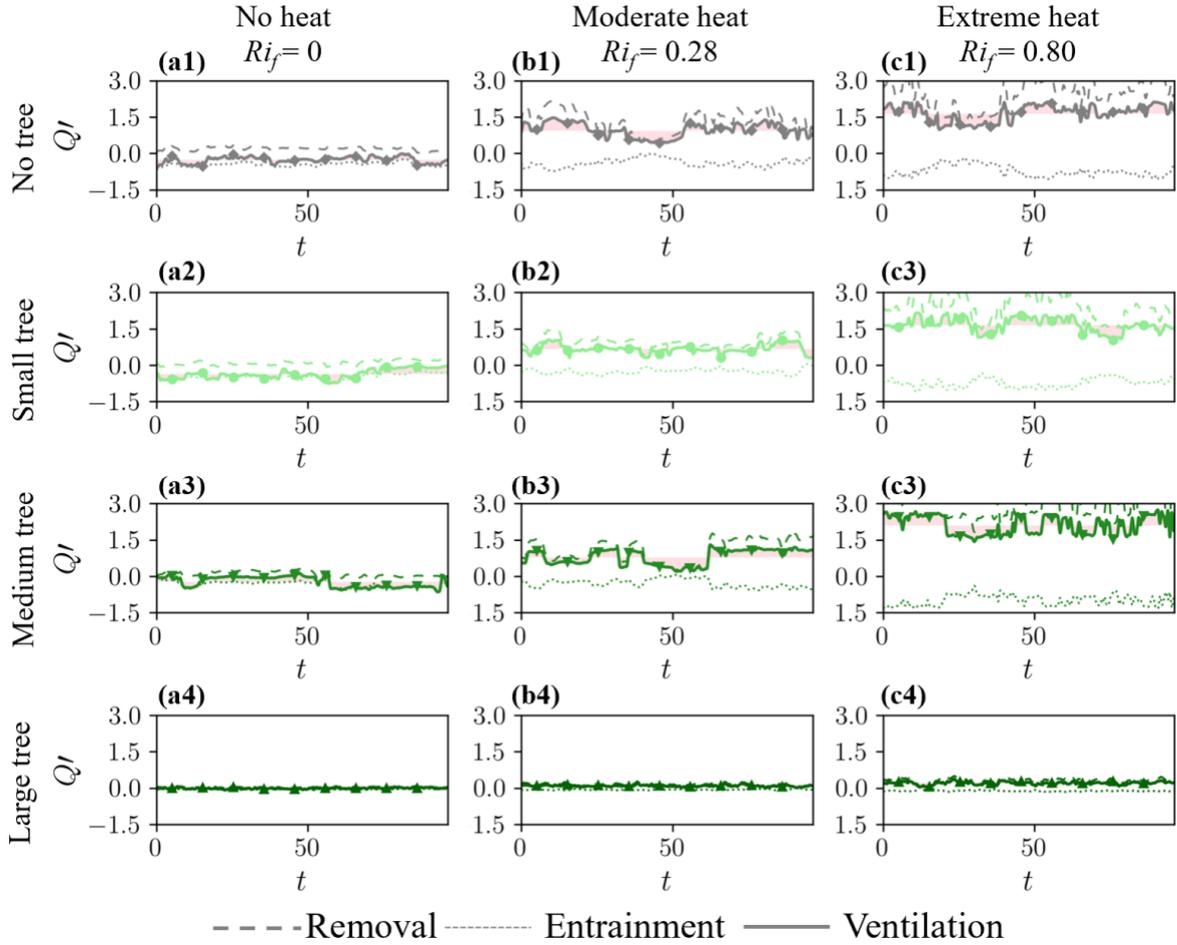

**Fig.5** Volumetric flow rate at the rooftop of the *flat canyon*. The solid lines represent timeseries of the volumetric ventilation rate ($Q'$), the dashed lines for removal, and the dotted lines for entrainment. Isothermal cases are shown in (a1) - (a4), moderate heat cases in (b1) - (b4), and extreme heat cases in (c1) - (c4). The shaded regions represent the turbulent ventilation rate ($Q'_t$).

In moderate and extreme heat, a positive ventilation rate implying air removal is seen in the canyon without a tree (Fig.5 b1, c1), which is due to the buoyancy-driven upward flow. For the cases with small and medium trees, the ventilation rate approximately remains at the same magnitude as the corresponding case without a tree, suggesting a negligible blockage effect of trees on buoyancy-driven air removal, as shown in Fig. 5b2-3, c2-3. However, for the canyon with large trees, the ventilation rate drops dramatically to a negligible magnitude, as shown in Fig. 5b4, c4. This is mainly due to the blockage by the tree canopy to the upward airflow.

A similar blockage effect due the presence of the trees is observed in the steep canyon. Large foliage reduces the dominant downward flush flow caused by the tall, downstream building, while in the meantime it suppresses the upward buoyancy-driven flow developed from the heated ground surfaces.

## 3.3 Momentum fluxes

To better understand why the roof-level ventilation is affected differently by the trees of different size subject to different heat conditions, quadrant analysis is performed with the airflow at the roof-level center of the canyon opening. The timeseries of four events, sweeps, ejections, outward interactions, and inward interactions are calculated for these cases without and with trees of different sizes and the results are shown in Fig.6. Since the airflow in the steep canyon is largely dominated by the downward flush flow (Zhao et al., 2021), the analysis is centred on the momentum fluxes in the flat canyon.

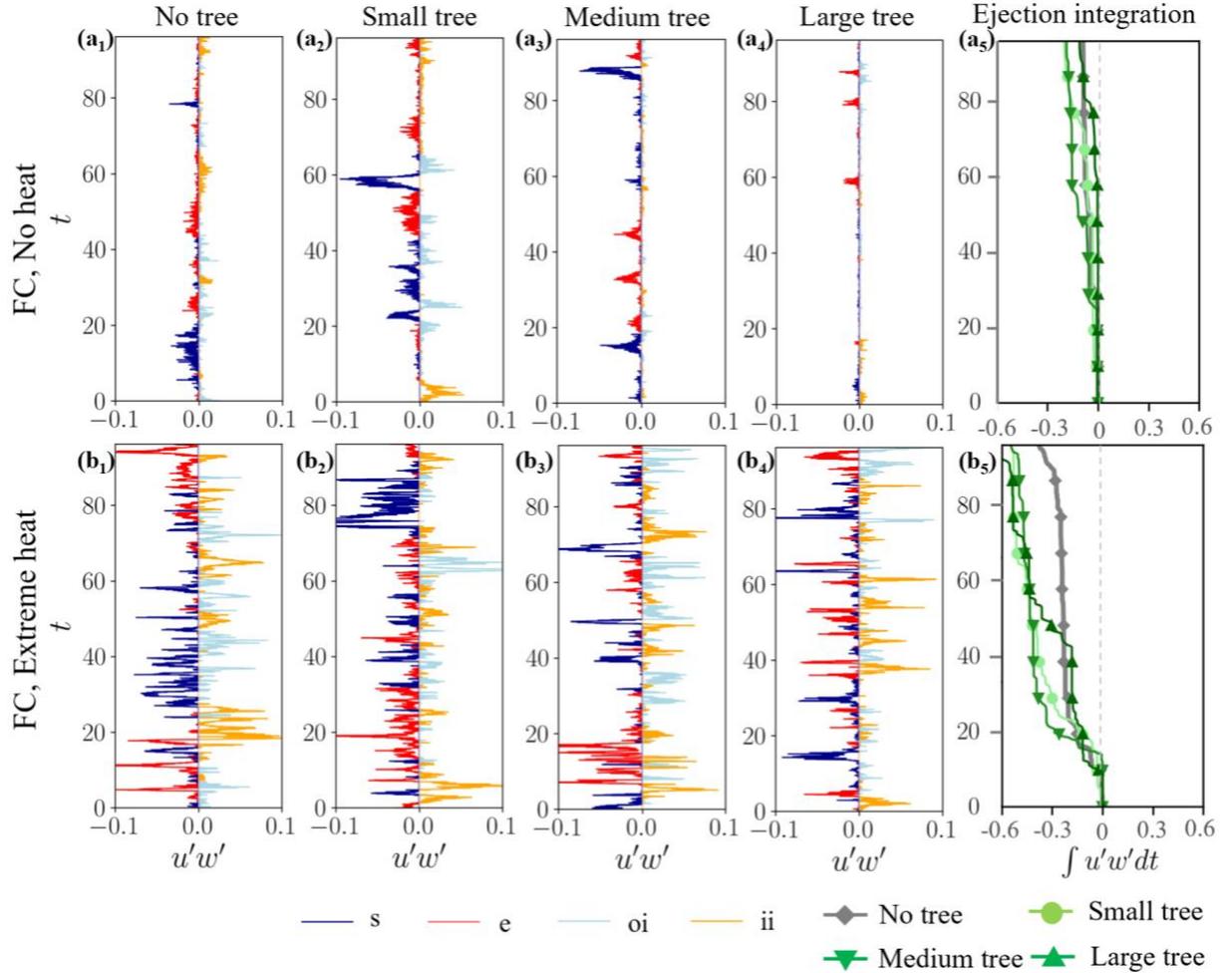

**Fig.6.** Time evolution of sweeps (s), ejections (e), outward interactions (oi), and inward interactions (io) obtained at the center of the roof level in the *flat canyon* with ($a_1$) no tree, ($a_2$) small trees, ($a_3$) medium trees, and ($a_4$) large trees in the isothermal condition; and ($b_1$) – ($b_4$) in extreme heat condition. Ejection integrated over time in ($a_5$) no heat and ($b_5$) extreme heat conditions. The nondimensional time series $t$ shown here is normalized by the time scale $t_r$.

For the flat canyon under isothermal conditions, the four events are all slightly amplified when small trees are present, as shown in Fig.6$a_2$, where spikes demote strong events. In the cases with medium and large trees, all four events dampen out gradually. Only minor ejections take place when large trees are present (Fig. 6a$_4$), which is consistent with the finding in the analysis of the volumetric ventilation rate in the case of large trees. The observation here suggests that small trees in street canyons may increase the roughness of a canyon ground surface and therefore enhance turbulent flow fluctuations at roof level. However, when the trees are large enough, such as comparable to the canyon height, the flow is less intermittent (stabilized) because the trees fill up a considerable portion of the canyon and the canyon-wide roughness is reduced to a certain extent. Fig.6a$_5$ shows the integration of the ejection over time, which clearly implies that the presence of the large trees stabilizes the flow significantly, whilst the presence of the small and medium tree facilitates airflow exchange with at the rooftop compared to the case without tree.

For the extreme heat case, the impact of trees on the four types of events taking place at the same location is clearly different compared to the isothermal cases as shown in Fig.6b$_1$-b$_4$. For the cases with medium and large trees, the suppression of the four types of events is not seen. Instead, the four types of events are all amplified significantly, since the airflow at the observation location is largely dominated by buoyancy and the fluctuating components of airflow velocities in all four cases are less affected by the tree size. This is also evident from the integration of the ejection overtime, as shown in Fig.6b$_5$.

To understand the spatial distribution of the sweeps and ejections, Fig.7 shows the instantaneous contours for two types of representative events in the isothermal and extreme heat case with large trees. Extreme ejection events appear around and above trees in the extreme heat case, which are not seen in the isothermal case. This is because the trees absorb heat and in turn, act as the heat source in the canyon. The strong fluctuating velocity component in the vertical direction alongside the negative streamwise velocity component induced by the tree leaves leads to massive ejection events.

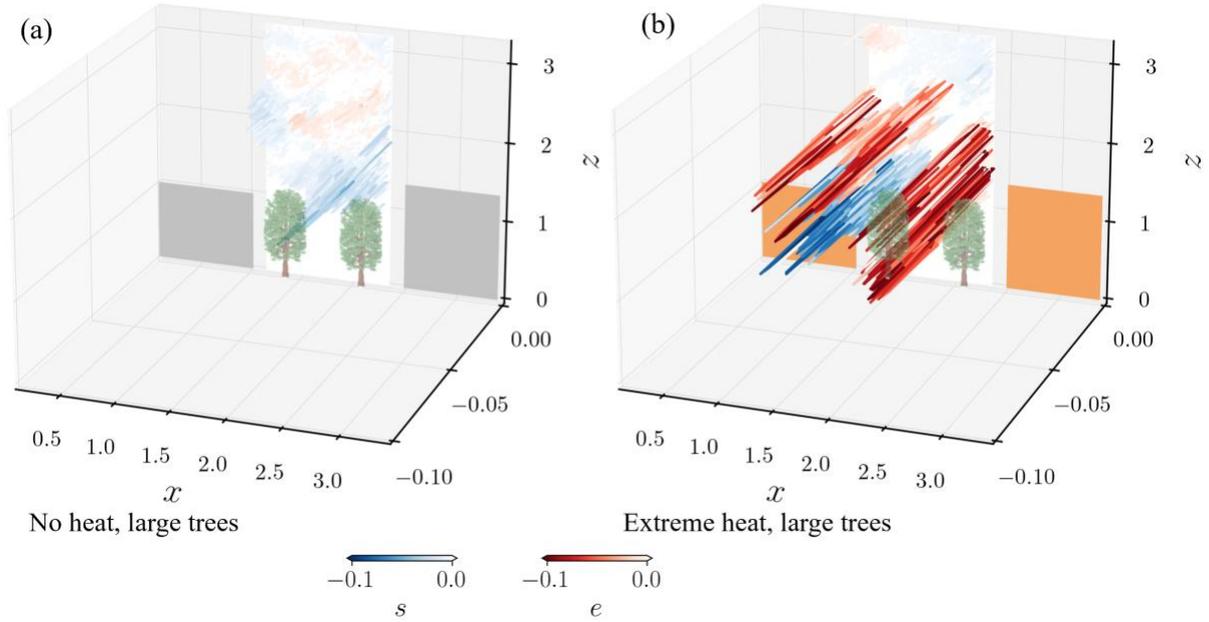

**Fig.7.** Representative instantaneous sweeps and ejections for: (a) isothermal condition with large trees, and (b) extreme heat with large trees.

### 3.4 Impact on air temperature at pedestrian-level

The air temperature at pedestrian level (1.8 m at full-scale) is analysed, as this information may be interesting for investigating pedestrian thermal comfort. It is noted that the impact of trees on air temperature of the region at pedestrian level may differ from the region above the tree canopy. This is because the airflow is largely decelerated below the tree canopy alongside the generation of vortices, while the airflow above the tree canopy is primarily dominated by the inflow. Fig. 8 shows the time series and time-average air temperature at pedestrian-level for the canyons without and with trees of different sizes. Here the canyons without trees are considered as the reference case. For the moderate heat cases, the effects of trees on heat removal from street canyons are different for the flat and steep canyons. It is seen in Fig.8a$_1$-a$_4$ that the pedestrian-level air temperature is apparently higher in flat canyons with trees compared to the other canyons. The higher pedestrian-level air temperature can be explained by the heat absorption by tree leaves and the blockage of the airflow due to the crown of the trees. The pedestrian-level air temperatures in the steep canyon with trees are slightly lower than that in the canyon without a tree. This is likely because the airflow mixing in the lower part of the canyon is enhanced by the downward airflow along the windward surface and the presence of porous trees' foliage, which enhances heat removal from the steep canyons.

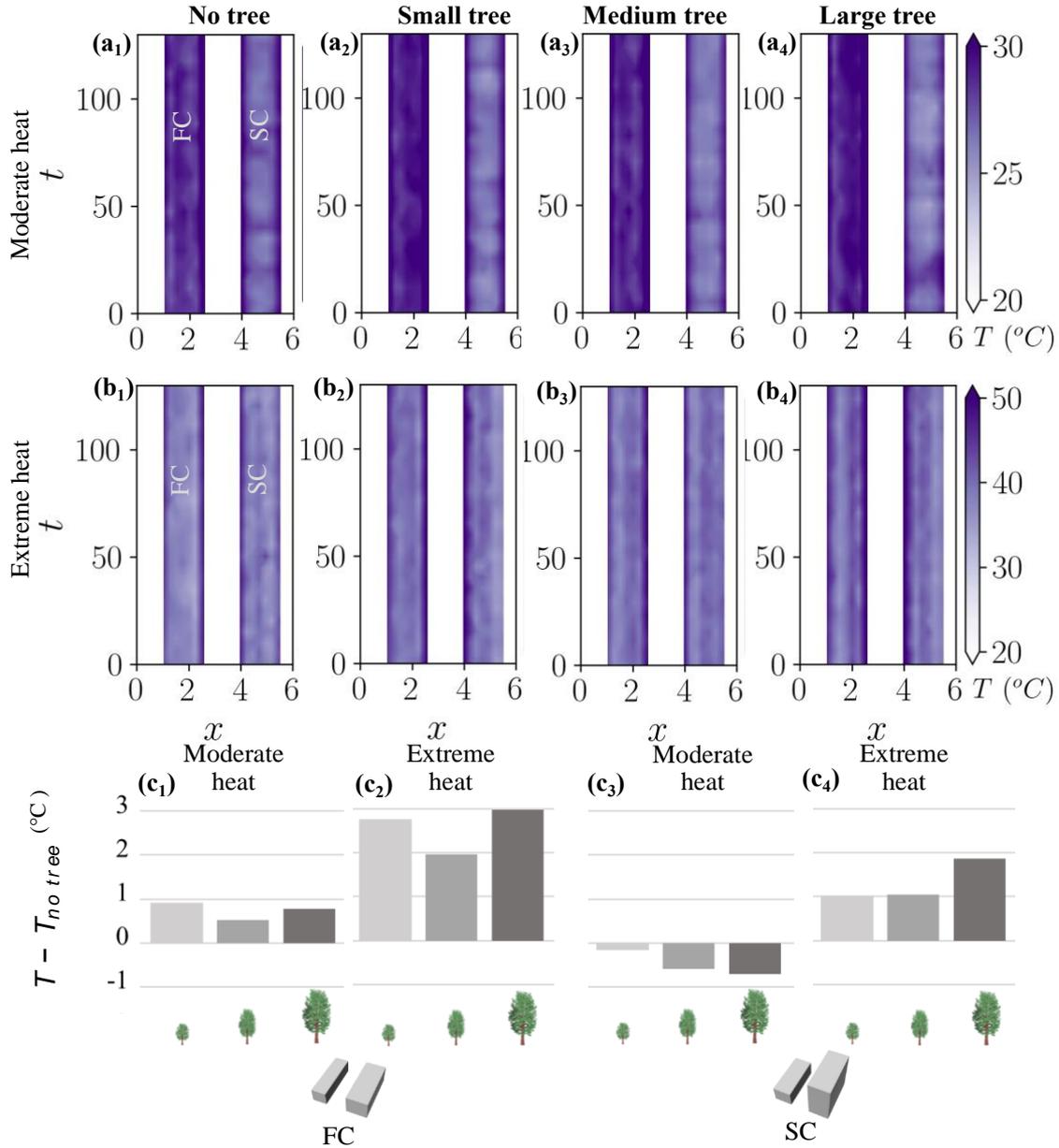

**Fig.8.** ($a_1$-$a_4$) and ($b_1$-$b_4$): Simultaneous time series of air temperature at pedestrian-level in the *flat* (FC) and *steep* (FC) canyons for moderate heat without ($a_1$) and with trees ($a_2$-$a_4$) of different sizes and for extreme heat without ($b_1$) and with trees ($b_2$-$b_4$) of different sizes. Time-average of air temperature of flat canyon subjected to moderate ($c_1$) and extreme ($c_2$) heat; steep canyon subjected to moderate ($c_3$) and extreme ($c_4$) heat.

The differences in time-average pedestrian-level air temperatures for cases without and with trees are shown in Fig. 8$c_1$-$c_4$. It is seen that in the flat canyon the presence of trees leads to a rise in pedestrian-level air temperatures by 0.5-1°C as a result of reduced heat removal, which corresponds to a temperature rise of 0.15 -0.3°C in full-scale settings at a wind speed of 4.2 m/s. This temperature rise due to the trees is not seen in the steep canyon, where instead the air temperatures are lowered slightly due to enhanced airflow mixing brough by the downward flush flow by the high-rise building.

For the cases subjected to extreme heat, higher air temperatures are observed at pedestrian level in both the flat and steep canyons due to the presence of trees, as shown in Fig.8$b_1$-$b_4$. Under this extreme heat condition, the airflow in the two canyons is buoyancy-driven and the trees predominantly act as porous media absorbing heat, at the same time weakening upward airflow from ground level. The temperature rise is up to 2-3°C in the flat canyon and 1-1.9°C in the steep canyon in measurements (Fig.8$c_1$, $c_4$), which corresponds to a rise of 0.6-1.5 °C and 0.3-0.6 °C in the two canyons in full-scale settings. These observations are made for a wind tunnel setup not including shading, as will be further addressed in the limitation section.

**3.5 Heat storage by leaves**

In the current experimental setup, trees in street canyons also absorb heat and may influence convective heat transfer between urban surfaces and ambient air. As discussed, this setup may resemble conditions during nighttime when tree transpiration does not function or is reduced substantially (Oren et al., 1999) . Fig.9 shows the air temperature in the canyon, the temperature of tree leaves and the air-leaf temperature difference for different tree size.

It is seen that the air temperature above rooftop level in both canyons is lower for moderate and extreme heat conditions when trees are present due to a reduction of heat removal from the canyons to the roof level, refer to Fig.9$a_1$-$a_4$ and Fig.9$b_1$-$b_4$. This is due to the fact that trees absorb substantial heat while reducing convective heat transfer from the ground and building surfaces to the roof-level approaching flow due to roughness of the leaves and blockage effect by tree crown.

The surface leaf temperature is obtained from infrared camera measurements, as shown in Fig.9$c_1$-$c_3$, $d_1$-$d_3$. Fig.9$c_4$ shows that under moderate heat condition, the temperature difference is around 8-10 °C, which corresponds to 2.5-3 °C in full-scale settings, estimated from the similarity of Richardson number determined based on the length scale of tree height and leaf-air temperature difference. Under extreme heat conditions, the maximum temperature difference is about 45 °C, which represents a 13.8 °C leaf-air temperature difference for the large trees. Such differences have been observed for tropical trees (Fauset et al., 2018).

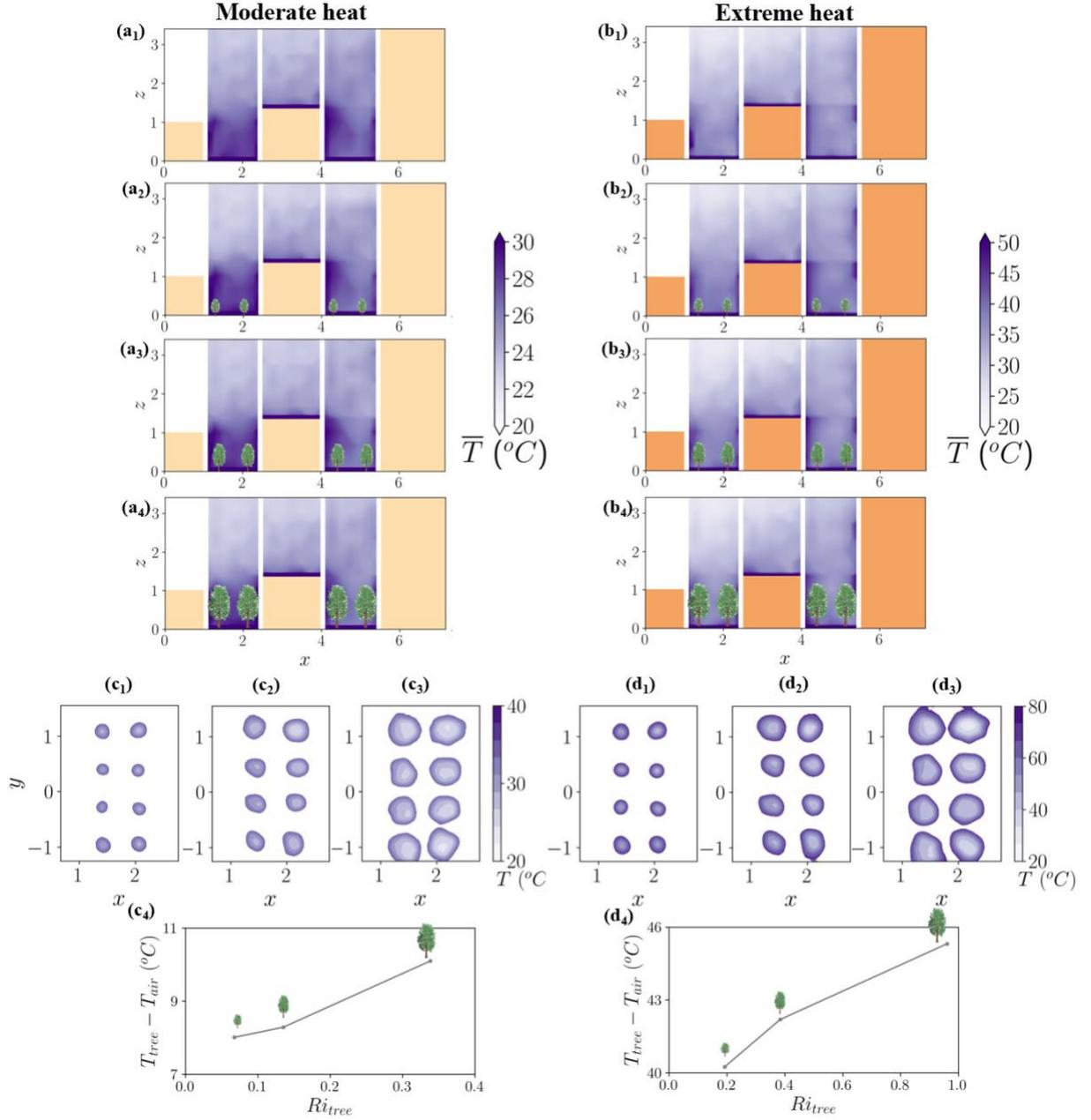

**Fig.9.** Temperature field in street canyons for cases without and with trees of different sizes exposed to (a$_1$-a$_4$) moderate and (b$_1$-b$_4$) extreme heat. Leaf surface average temperatures (c$_1$-c$_3$, d$_1$-d$_3$) and temperature difference between tree and approaching airflow for (c$_4$) moderate and (d$_4$) extreme heat conditions.

## 4 Discussion
### 4.1 Dynamics of flow structures

The measurement results reveal that trees decelerate roof-level incoming boundary layer flow, as heat removal barrier above pedestrian level, as a ventilation barrier at roof level, and as heat storage material absorbing substantial heat from the surrounding environment, all depending on the size of the trees, buoyancy condition of the airflow in the canyons, and street canyon morphologies among others.

The airflow dynamics observed in the measurements is fundamentally altered by the presence of trees and extreme heat from building and ground surfaces, as illustrated in Fig.10. The flow structures are illustrated by streamlines of time-averaged flow field observations from the PIV measurements. The presence of multiple large trees introduces significant mutual sheltering effects in the urban canopy, which reduces the downward flow along the windward surface, resulting in that the canyon-wide vortical flows do not establish as those in canyons without trees, as illustrated in Fig.10a-b.

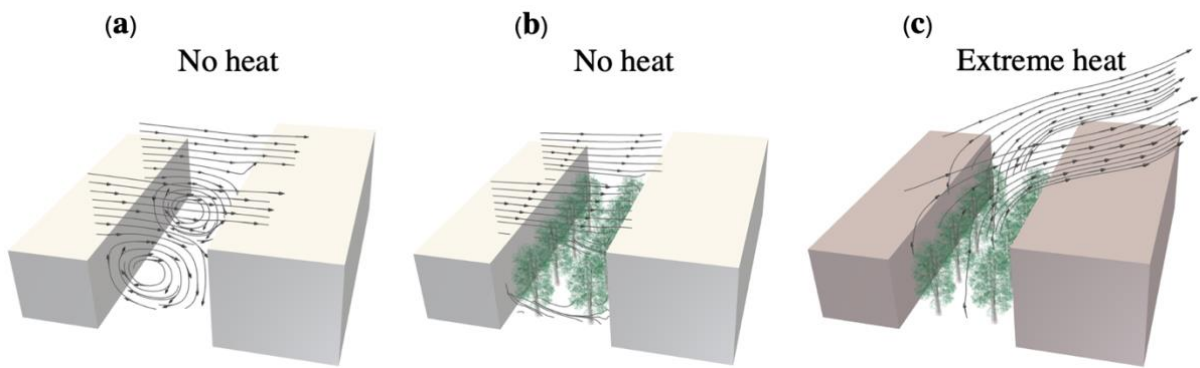

**Fig.10.** The observed dominant flow structures: (a) without trees in the isothermal condition, (b) with large trees in the isothermal condition, and (c) with large trees in the extreme heat condition.

In the extreme heat condition, the buoyancy effect plays a dominant role and results in strong upward flow leaving the canyon, which could be beneficial for heat removal from street canyon to rooftop as illustrated in Fig.10b-c. In this condition, building and ground surfaces and even the trees act as heat source to heat up the airflow.

The studied cases could resemble conditions in street canyons during night. However, the presence of trees may reduce the sky view factor of the canyon and thus reduce longwave radiation and heat removal from streets at nighttime. The airflow dynamics also depends on the spacing of trees and the foliage cover relative to the ground surface of a canyon. Parametric studies are needed to reveal the joint influences of these factors.

## 4.2 Limitations of the present work

The present experimental study focuses on the aerodynamic and thermal storage effects of trees of different sizes, which could resemble street canyon microclimate during nighttime. Shading and transpiration, which are two essential functions of trees during daytime are not modeled in the wind tunnel experiments. Further, the limitation of the present work lies in the limited parametric exploration of the influence of the porosity and leaf area density (LAD) of trees, spacing and street canyon cover by trees. The LAD may vary considerably among different species and the climate zones. The porosity range used in this study may represent trees without *dense leaves*, which may lead to an underestimation of drag compared with real trees having dense leaves.

The spacing of trees, in addition to the LAD of an individual tree, also affects the airflow dynamics in canyons, volumetric ventilation rate, pedestrian-level microclimate, etc. In the present study the ratio of foliage cover to canyon opening ranges from 25% to 3%. Further explorations concerning mutual sheltering effects by trees and a large range of the foliage cover are needed to complement the understanding obtained in this work.

Thermal storage of real fresh leaves is generally higher than that of the fiber-like synthetic wadding materials used in this experimental modelling. The specific heat capacity of fresh leaves varies from 1255 to 5174 J kg$^{-1}$ K in 6 types of leaves measured (Jayalakshmy et al., 2010), and the specific heat capacity of fiber-like synthetic wadding materials is approximately 1500 J kg$^{-1}$ K. Further studies with different thermal storage by leaves is necessary.

Street canyon microclimate is in fact dominated by multiscale phenomena and complex involving different physical phenomena. Well-controlled wind tunnel measurements of shading and transpiration in street canyons are much needed for numerical model validations. The LAD and their optical porosity in measurements can be well characterized for numerical settings. Future experimental studies should be designed involving the transpiration rate of real model trees involving the control of soil moisture content which is an essential factor that has been mostly assumed to be constant in the literature.

## 5    Concluding remarks

The impact of the size of trees, when trees show different ages, on street canyon microclimate is studied experimentally in the ETHZ/Empa Atmospheric Boundary Layer wind tunnel using planar PIV, airflow temperature field and leaf surface temperature measurements. Trees characterized by a range of relative heights ($H_{rh} = H_{tr}/H_{ca}$) are considered in scaled-down realistic street canyons that are exposed to isothermal (no heat), moderate, and extreme heat conditions, as explained in section 2. Depending on the canopy cover of trees, the intensity of heat in street canyons, street canyon morphologies, and trees may play different roles.

When trees are present in the flat canyon in the isothermal condition, the boundary layer flow approaching above street canyons can be decelerated by trees, as seen from the significant damping of sweeps, ejections, outward interactions and inward interactions as stated in section 3.3. The ejection integrated over time, denoting momentum transport in the streamwise and upward direction, is dampened from $\int u'w'dt = -0.2$ (at $t$ of 80) in the case with large trees ($H_{rh}$ = 1.10) to a negligible magnitude in the case with small trees ($H_{rh}$ = 0.43).

As trees grow beyond the center of the canyon height (e.g., $H_{rh}$ = 0.70 for the flat canyon), the shear flow cannot drive anymore a canyon-wide vortex circulation in the isothermal cases because the flow is significantly decelerated by the tree leaves (section 3.1). For the same reason, the large trees reaching above the canyon (e.g., $H_{rh}$ = 1.10 for the flat canyon) dramatically lower the street canyon ventilation (i.e., air exchange) at the roof level (section 3.1).

When the street canyons are exposed to heat, the presence of trees reduced ventilation from pedestrian level to rooftop airflow may result in the air temperature rise in the canyons. The rooftop average air removal ($Q' > 0$) is approximately reduced from 1.75 to 0.30 as trees grow from a small relative height ($H_{rh}$ = 0.43, flat canyon) to a large relative height ($H_{rh}$ = 1.10, flat canyon), as shown in section 3.2. This ventilation reduction is largely due to the deceleration of buoyancy-driven upward mean flow in the canyon. The temperature increase ranging from 0.15 °C to 1 °C is observed in the flat canyon exposed to extreme heat. However, as an exception, trees are found to also slightly lower pedestrian level air temperature in the steep canyon for moderate heat conditions, which is very likely due to enhanced thermal mixing in the lower part of the steep canyon.

It is worth noting that the given experiments do not consider the transpiration of leaves and the longwave radiation to a cold sky during the night. The heat storage capacity of the fiber-like synthetic wadding material is also lower than that of fresh leaves. These suggest that the temperature increase observed in the present case with artificial leaves could be overestimated to some extent. The present setup is not representative for studying the impact of street trees during the daytime when transpirative cooling from leaves and shading by trees play important roles. To study these mechanisms, living vegetation subject to a series of realistic humidity conditions and a solar simulator may be adopted, where the impact of wind on transpirative cooling and canyon temperature distribution is worth investigating. In addition, the joint effects of the tree-to-canyon foliage cover and relative height, canyon geometry, different heat intensities in canyons, and varied wind conditions deserve future parametric studies.

**Acknowledgments**
We thank Beat Margelisch, Roger Vonbank and Claudio Mucignat for preparing the measurement equipment.

**Appendix A.** Configuration of trees at full scale and wind tunnel scale. T-T spacing denotes the distance between the trees from trunk to trunk in the spanwise direction; T-W spacing denotes the distance from a tree's trunk to its nearest building wall.

| Tree's | Full scale | | | | Wind tunnel scale | | | | | |
|---|---|---|---|---|---|---|---|---|---|---|
| | Height (m) | Crown (m) | T-T Spacing (m) | T-W Spacing (m) | Height (mm) | Crown (mm) | T-T Spacing (mm) | T-W Spacing (mm) | Porosity (%) | Foliage cover (%) |
| Large (L) | 22.4 | 9.6 | 15 | 5 | 140 | 60 | 94 | 31 | 72-80 | 25 |
| Medium (M) | 14.4 | 6.4 | 15 | 5 | 90 | 40 | 94 | 31 | 72-80 | 11 |
| Small (S) | 8.8 | 4 | 15 | 5 | 55 | 25 | 94 | 31 | 72-80 | 4 |